\documentclass[pra,twocolumn]{revtex4-1}
\usepackage{graphicx}
\usepackage{amsmath}
\usepackage{color}
\usepackage{comment}

\begin{document}

\title{Adiabatic potentials using multiple radio frequencies}

\author{T.~Morgan}
\email{tmorgan@phys.ucc.ie}
\affiliation{Quantum Systems Unit, Okinawa Institute of Science and Technology, Okinawa, Japan}
\affiliation{Department of Physics, National University of Ireland, UCC,
  Cork, Ireland}

\author{Th.~Busch}
\affiliation{Quantum Systems Unit, Okinawa Institute of Science and Technology, Okinawa, Japan}
\affiliation{Department of Physics, National University of Ireland, UCC,
  Cork, Ireland}

\author{T.~Fernholz}
\affiliation{School of Physics \& Astronomy, The University of
Nottingham, University Park, Nottingham, NG7 2RD, UK}

\begin{abstract}
  Adiabatic radio frequency (RF) potentials are powerful tools for creating advanced trapping geometries for ultra-cold atoms. While the basic theory of RF trapping is well understood, studies of more complicated setups involving multiple resonant frequencies in the limit where their effects cannot be treated independently are rare. Here we present an approach based on Floquet theory and show that it offers significant corrections to existing models when two RF frequencies are near degenerate. Furthermore it has no restrictions on the dimension, the number of frequencies or the orientation of the RF fields. We show that the added degrees of freedom can, for example, be used to create a potential that allows for easy creation of ring vortex solitons.
\end{abstract}

\maketitle

\section{Introduction}
\label{sec:Introduction}

Coupling resonant RF radiation to the spins of atoms in
inhomogeneous magnetic fields allows to locally address clouds of
ultracold atoms with high spatial resolution. This technique is
commonly used in combination with atom chips, where it
offers huge flexibility in the types of potentials that can be
created \cite{Folman:02, Fortagh:07}. Since the RF
fields can also be time dependent, this technique offers the possibility to control the atoms dynamically and has, for
example, been used to split Bose-Einstein condensates (BECs)
\cite{Schumm:05}. By today significant work has been done in the 
modelling of RF systems to create highly non trivial trapping
geometries \cite{Lesanovsky:06,Fernholz:07}, including annular
traps \cite{Morizot:06}, examining effects beyond the rotating wave
approximation \cite{Hofferberth:07} and investigating non-linear
Zeeman effects \cite{Sinuco:12}.

The use of multi-frequency fields has recently been suggested to
create multi-well or even periodic potentials
\cite{Courteille:06,Morgan:11}. These can also be time
dependent and in \cite{Morgan:11} a method to coherently control
the centre of mass motion of both, a single atom and a cloud of
interacting atoms, was suggested. While this work highlighted how
versatile multi-frequency RF potentials are for experimentally
realistic parameters, the theoretical model used could only describe limited tunnel coupling strengths between two potential wells, as it required all radio frequencies to be well separated. For large tunnel couplings, however, it is usually necessary to have potentials where the curvature can be controlled over short distances, which, using RF technology, requires two frequencies to be brought close. In such situations the effects of both frequencies have to be considered simultaneously and a more advanced model is necessary.

In this work we address the question of how to accurately deal with near degenerate frequencies  and take full advantage of all degrees of freedom offered by RF fields by using Floquet theory and comparing the results to the currently most common model \cite{Courteille:06}. Floquet theory
allows to describe the interaction between a periodically
oscillating field and a quantum system and its basic idea is to
replace the semiclassical time-dependent Hamiltonian with a
time-independent Hamiltonian of infinite dimensions
\cite{Shirley:63}. We will show that this method allows to obtain highly accurate potentials.

Floquet theory is applicable to single and many-mode systems \cite{Ho:83} and it has been extensively used in
studying multi-photon processes for atoms in strong laser fields
\cite{Ho:84}. For example Drese an Holthaus \cite{Drese:99} used Floquet theory  to
describe adiabatic potentials produced by short laser pulses in
order to stimulate a STIRAP (Stimulated Raman Adiabatic Passage)
process. More recently this was extended to study multiphoton resonance dynamics driven by
intense frequency-comb laser fields using many-mode Floquet
theory \cite{Son:08}. A
comprehensive overview of the application of Floquet theory to
atomic and molecular multiphoton processes in intense laser fields
can be found in \cite{Chu:04}.

The layout of this paper is as follows. In Sec.~\ref{sec:GenRF} we briefly review the solutions of the  Hamiltonian describing a magnetically trapped atom, which is exposed to a single-frequency RF field. In Sec.~\ref{sec:PiecewiseResonance} we discuss the piecewise resonance model for dealing with multiple RF fields, which was developed in
\cite{Courteille:06}, and show how it breaks down when two frequencies come too close. In Sec.~\ref{sec:FloquetModel} we show how Floquet
theory can be used to obtain a more accurate description and in
Sec.~\ref{sec:PotentialGen} we compare the two approaches and discuss the corrections and benefits of the Floquet approach. To demonstrate the power of the approach, we detail in
Sec.~\ref{sec:2dPotential} how an extension to higher dimensions
can be done and show how controlling the orientation of the RF
field can be used to produce non trivial trapping geometries. As an
example, we show how controlling the tunnel coupling in multidimensional RF traps can allow to create ring vortex solitons in Bose-Einstein condensates. Finally we conclude in Sec.~\ref{sec:Conclusion}.

\section{Radio-frequency dressed traps}
\label{sec:GenRF} RF trapping is based on the idea of coupling
magnetic sublevels in the presence of an inhomogeneous magnetic
field. Consider a hyperfine atomic groundstate manifold with total
spin $F$. In the presence of a static magnetic field of modulus
$B_s$, the Zeeman sublevels, $m_F=-F,...,F$ undergo a linear
splitting according to $E_{m_F}=m_F g_F\mu_B B_s$, as long as the
field strength is not too large and second order splitting is
negligible. Here, $g_F$ is the atomic g-factor and $\mu_B$ is the
Bohr magneton. Irradiating such a system with a radio frequency
${\bf B}_\text{RF}(\omega t)$, which can excite atomic Larmor
precession, will couple the sublevels $|F,m_F\rangle
\leftrightarrow |F,m_F\pm 1\rangle$ with spatial selectivity due
to the inhomogeneity of the magnetic field and the spatially varying
Larmor frequency.

For simplicity, let us first consider a one-dimensional situation with
a static field pointing and increasing linearly along the
$z$-direction, i.e. ${\bf B}_s(z)=B_s(z){\bf e}_z=G z{\bf e}_z$ where G is the magnetic field
gradient. Such a situation arises, for example, when an atom
travels along an axis of a quadrupole field. It is convenient to
express the oscillating field in complex spherical components
($\sigma^+$, $\sigma^-$, and $\pi$-polarizations), such that in Cartesian coordinates
\begin{equation}
  \label{eq:rfBdef}
  {\bf B}_\text{RF}(\omega t)=\text{Re}\left[\left(\begin{array}{c}
  \frac{1}{\sqrt{2}}(B_++B_-)\\
  \frac{i}{\sqrt{2}}(B_+-B_-)\\
  B_{\pi}
  \end{array}\right)\cdot e^{-i\omega t}\right]\; .
\end{equation}
The total Hamiltonian for the magnetic interaction can then be
written as
\begin{equation}
\label{eq:Hamiltonian} H(z,t)=\mu_B g_F{\bf \hat{F}}\cdot({\bf
B}_s(z)+{\bf B}_\text{RF}(\omega t))\;,
\end{equation}
where the angular momentum operator is expressed in units of $\hbar$.
Transforming to a frame rotating about the $z$-axis at frequency $\omega$ using
$H_\text{rot}=i\hbar\dot{U}U^{-1}+UHU^{-1}$ with $U=e^{i F_z
\omega t}$ leads to $H_\text{rot}=\mu_B g_F{\bf \hat{F}}\cdot{\bf
B}_\text{eff}$, where the effective field in Cartesian coordinates
is given by
\begin{equation}
  \label{eq:rfBeff}
  {\bf B}_\text{eff}=\left(\begin{array}{c}
  \frac{1}{\sqrt{2}}\text{Re}[B_-e^{-i2\omega t}]+\frac{1}{\sqrt{2}}\text{Re}[B_+]\\
  \frac{1}{\sqrt{2}}\text{Im}[B_-e^{-i2\omega t}]-\frac{1}{\sqrt{2}}\text{Im}[B_+]\\
  \text{Re}[B_\pi e^{-i\omega t}]+B_s(z)-\frac{\hbar\omega}{\mu_B g_F}
  \end{array}\right).
\end{equation}
Expressing this Hamiltonian in the spherical basis ${\bf e}_{\pm}=({\bf e}_{x}\mp i{\bf e}_{y})/\sqrt{2}, {\bf e}_{\pi}={\bf e}_{z}$ then results in
\begin{align}
\label{eq:SphericalHamiltonian}
  H_\text{rot}=\frac{\mu_B g_F}{2}[&(B_s(z)-\frac{\hbar\omega}{\mu_Bg_F})\hat{F}_z+\frac{B_+}{\sqrt{2}}\hat{F}_+\\
   &+\frac{B_-}{\sqrt{2}}\hat{F}_-e^{-i2\omega t}+B_{\pi}\hat{F}_ze^{-i\omega t}]+c.c.\nonumber\;,
\end{align}
with the usual definitions $\hat{F}_{\pm}=(\hat{F}_x\pm i
\hat{F}_y)$.

If the RF field is purely $\sigma^+$-polarized, the effective
field and rotating frame Hamiltonian are time-independent. In 
this case, the situation in the rotating frame is fully equivalent 
to an atom moving in an inhomogeneous field across a non-zero field
minimum, defined by the resonance condition $\mu_B g_F B_s(z) = \hbar\omega$.
Conversely, the widely used Ioffe-Pritchard type trap, 
where the field-zero of a 2D quadrupole field is lifted with an 
orthogonal applied field $B_I$, can be viewed as a dressed state trap 
with zero coupling frequency. The quasi-energies, i.e., the eigenvalues 
of the Hamiltonian in the rotating frame can be obtained straightforwardly as
\begin{align}
    \label{eq:SingleRotExactE}
    E_{m_F} &= m_F \mu_B g_F B_\text{eff}\\
    \nonumber &=m_F \mu_B g_F
    \sqrt{\left( B_s(z)-\frac{\hbar\omega}{\mu_B g_F }\right)^2+ \frac{\left|B_+\right|^2}{2} }\; ,
\end{align}
where the $m_F$ are eigenvalues of the spin component aligned with
the effective field, i.e., rotating in the laboratory frame.

\begin{figure}[tb]
\includegraphics[width=1\linewidth]{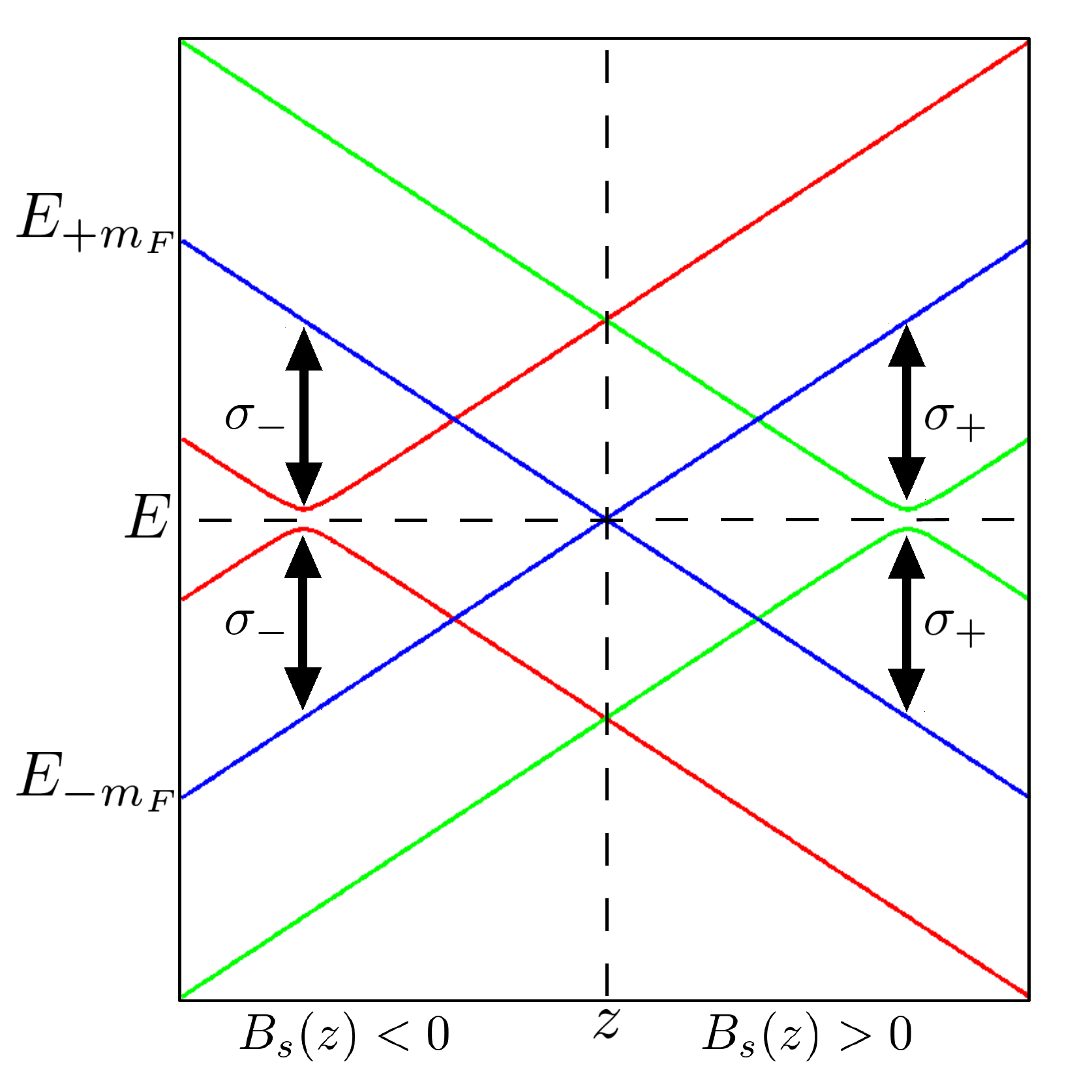}
\caption{(Color online) A schematic diagram of the eigenvalues of the Hamiltonian
in the rotating frame for both, $\sigma_+$ (green lines) and
$\sigma_-$ (red lines) polarised RF fields irradiating an atom in
a linear static field $B_s(z)=G z$. The blue lines show the bare
Hamiltonian eigenvalues $E_{+m_F}$ and $E_{-m_F}$. The dashed
black lines indicate $z=0$ and $E=0$.} \label{fig:SCHEME}
\end{figure}

If polarization components other than $\sigma^+$ are present, one
can apply the rotating wave approximation (RWA) by discarding the
fast oscillating terms as long as $|\hbar\omega|\gg
g_F\mu_BB_{\text{eff}}$ \cite{Tannoudji:92}. This however, will neglect shifts in the energy spectrum, which are known as
Bloch-Siegert shifts \cite{Bloch:40}. A second order correction is
caused by a $\sigma^-$ field component, which can be seen as
equivalent to a $\sigma^+$ polarized field at negative frequency, leading to off-resonant coupling between states with $\Delta
m_F=\pm1$ (see Fig.~\ref{fig:SCHEME}). Other higher order effects
can also stem from $\pi$ polarised components.

\section{Piecewise Resonance Model}
\label{sec:PiecewiseResonance}

More control and complexity can be introduced into RF dressed
potentials by using multiple frequencies $\omega_n$. Using the
method outlined above, one can in principle apply an iterative
procedure by performing consecutive rotating wave approximations
for each frequency. However, it is then necessary at each step to
transform all oscillating fields into the new rotating frame and
determine new spherical polarization components with respect to the
effective field of the previous step. This procedure will lead to
an accumulation of errors and can suffer from a certain
arbitrariness as the order of eliminating individual frequencies
is not defined.

However, if the individual frequencies are spaced sufficiently far apart and the coupling is small with respect to the detuning, one can approximate the dynamics locally by considering only the nearest resonance frequency, $\omega(z)=\omega_{n}$. Formally this means that $n$ is chosen such that $\lbrack \mu_B g_F B_s(z) - \hbar \omega_{n} \rbrack$ is minimized at any position $z$. In the language of the above consecutive approach, this is equivalent to approximating the effective field by its $z$-component before decomposing the polarisation components for the next step. In the following, we briefly revisit the method outlined in \cite{Courteille:06} to analyse the resulting potentials.

With the local frequency approximation, and after applying the RWA, the Hamiltonian of Eq.~(\ref{eq:SphericalHamiltonian}) for a system of total spin $F=\frac{1}{2}$ can be expressed as
\begin{equation}
  \label{eq:rfHamiltonian}
  H(z)=\frac{1}{2}
    \left(\begin{array}{cccc}
    \mu_B g_F B_s(z) - \hbar \omega_n &
    \hbar \Omega_n  \\
    \hbar \Omega^*_n &
    \hbar \omega_n-\mu_B g_F B_s(z)
          \end{array}\right) \;,
\end{equation}
where $\Omega_n=\frac{\mu_B g_F}{\hbar\sqrt{2}}B_{n_+}$ is the Rabi frequency of the $n^{\text{th}}$ RF field with $\sigma^+$-polarisation and amplitude $B_{n_+}$ (or $\sigma^-$-polarisation for negative $\omega_n$). For practical reasons, experiments are often performed with linearly polarised RF fields, orthogonal to the static field. In this case, the $\sigma$-components are given by $B_{n_{\pm}}=(B_{n_x}\mp iB_{n_y})/\sqrt{2}$. If only one linear polarisation is used, say $B_x$, each frequency produces two equally strong resonances at $\pm \omega_n$ or $\pm B_s$ with Rabi frequency $\Omega_n=\frac{\mu_B g_F}{2\hbar}B_{n_x}$.
 
\begin{figure}[tb]
\includegraphics[width=1\linewidth]{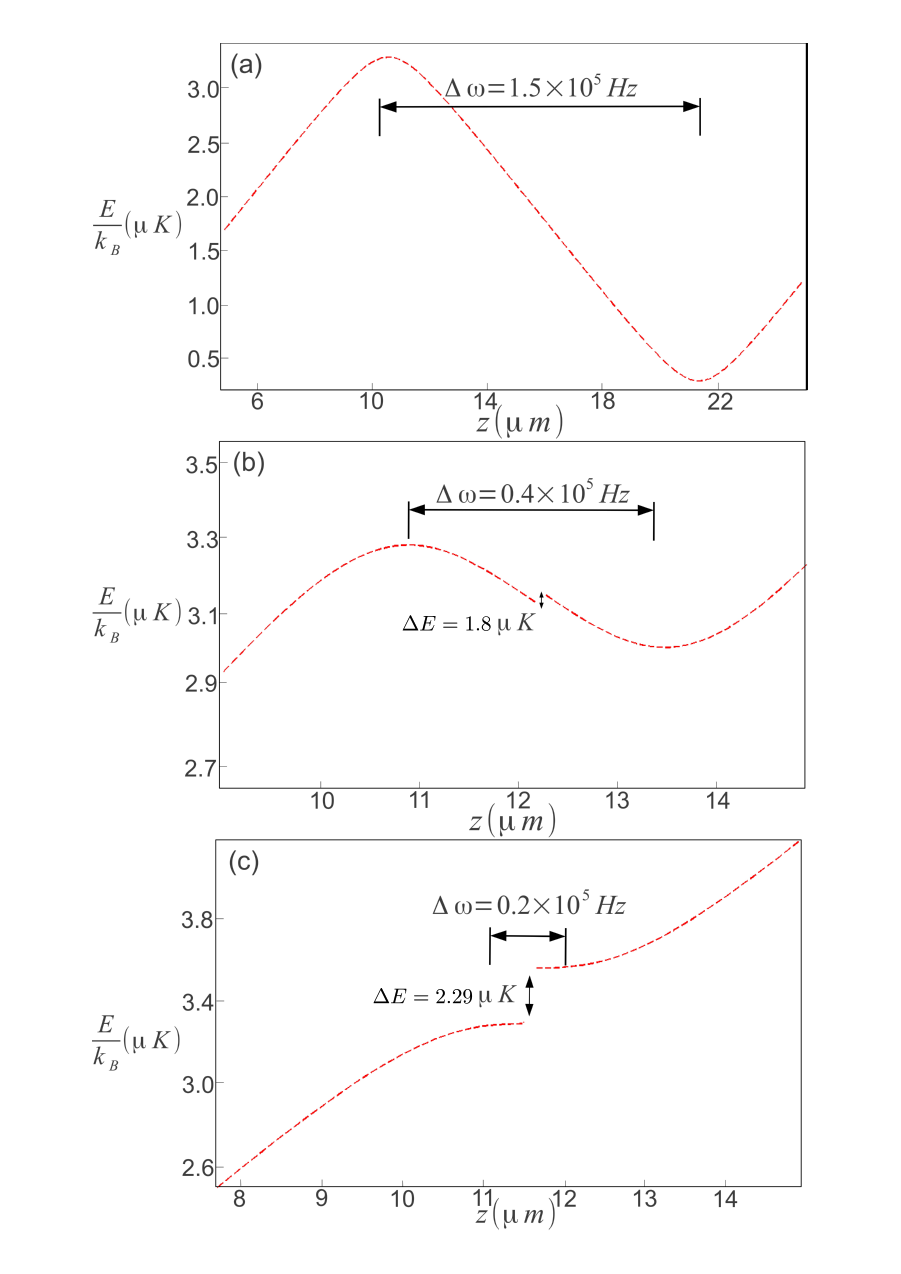}
\caption{(Color online) Adiabatic RF potentials for an atom trapped in a IP trap
  irradiated by two linearly polarised RF fields with
  different frequency separation $\Delta \omega$ for
  $\omega_1=1.5\times 10^5$ Hz and Rabi frequencies $\Omega_1=\Omega_2=9.274\times 10^{4}$ Hz. Note that the axis for each plot changes so 
  that the important features of each potential are resolved. In all cases $m_F=\frac{1}{2}$ and $g_F=1$.}
\label{fig:PWPotential}
\end{figure}

For large detuning the eigenenergies of the two-level RWA-Hamiltonian can be written as
\begin{align}
  E_{\pm}(z) &= \pm \frac{1}{2} \sqrt{ \hbar^2 |\Omega|^2
              + \lbrack \mu_B g_F B_s(z) - \hbar \omega_n \rbrack^2}\\
            &\approx \pm \frac{1}{2} \lbrack \mu_B g_F B_s(z)
                   - \hbar \omega_n \rbrack
                   \pm \frac{\hbar^2 |\Omega|^2}
                   {4 \lbrack \mu_B g_F B_s(z) - \hbar \omega_n \rbrack} \nonumber\;,
\end{align}
where the second step is valid far from the resonance, $\hbar\Omega \ll \lbrack \mu_B g_F B(z) - \hbar \omega_n \rbrack$, and the resulting second term can be viewed as a Stark shift. The effect of all other RF fields which are not closest to resonance can be approximated by creating an effective Stark shift, which is the sum of the shifts of all RF fields of frequency $\omega_j$ ($j \neq n$) \cite{Courteille:06},
\begin{equation}
  L_n(z)=\sum_{j\not=n} \frac{\hbar^2 |\Omega|^2}
         {4 \lbrack \mu_B g_F B_s(z) - \hbar \omega_{j} \rbrack} \; .
\end{equation}
This leads to a correction of the energies, which are now given by
\begin{equation}
  E_{\pm}(z) =\pm\frac{1}{2}\sqrt{ \hbar^2 |\Omega|^2 + \lbrack \mu_B g_F B_s(z)
              - \hbar \omega +2 L_n(z) \rbrack^2}\;.
\end{equation}
From this, and considering that the couplings are strong enough to
yield a Landau-Zener transition probability close to unity, the
resulting adiabatic potential can be written as
\begin{equation}
  V_{ad\pm}(z)=(-1)^{n} \left[ E_\pm(z) \mp \frac{\hbar \omega_{n}}{2} \right]
               \mp \sum^{n-1}_{k=1}(-1)^k \hbar \omega_k \;.
\end{equation}
In Fig.~\ref{fig:PWPotential} we show the adiabatic RF
potentials as seen by an atom in a static field gradient irradiated by two RF fields with frequencies, $\omega_1$ and $\omega_2$, for three
different values of $\Delta \omega=\omega_2-\omega_1$. In order to set scales applicable to typical atom chip experiments, all work presented here assumes a static field in the form of an Ioffe-Pritchard (IP) type trap with radial field profile
\begin{equation}
\label{eq:IP}
  B_s(r)=\sqrt{G^2 r^2 + B_I^2} \; ,
\end{equation}
where we chose a magnetic field gradient $G=1 T m^{-1}$ and an offset field $B_I=1 \mu T$.

While for well spaced frequencies the potential resulting from the above procedure is visually smooth (see Fig.~\ref{fig:PWPotential} (a)), a discontinuity appears for decreasing values of $\Delta \omega$, which grows as the frequencies approach each other. Treating the RF fields independently from each other is therefore no longer appropriate. In order to describe these situations we will explore the advantages a treatment using Floquet theory offers. Situations in which this becomes important are, for example, potentials in which the curvature changes on a small spatial scale.

\section{Floquet Model}
\label{sec:FloquetModel}

Floquet theory provides a practical and powerful nonperturbative method for studying the interaction of a quantum system with an oscillating field, such as ionization and multiphoton excitation processes. It allows to find the solutions of a Schr\"odinger equation with a temporally periodic Hamiltonian by representing it as an infinite matrix \cite{Shirley:63}.
By using a Floquet approach to describe a system containing multiple RF fields we can ensure that higher-order frequency shifts, which are neglected by the piecewise resonance model and other approaches which rely on the RWA, are accounted for.

As in the previous section, we aim to describe the state of an atom confined to a static
magnetic field ${\bf B}_s({\bf r})$ subjected to a radio-frequency field ${\bf B}_{RF}(\omega t)$ with N different frequencies. However, unlike the previous section, we consider linearly polarised fields that can have arbitrary orientation relative to the static field by defining the $\sigma_{+}$ and $\sigma_{-}$ components of the RF field as 
\begin{subequations}
\begin{align}
B_{+}&=\frac{\alpha \cos(\theta)- i \alpha \sin(\theta)}{\sqrt{2}}\;,\\
B_{-}&=\frac{\alpha \cos(\theta)+ i \alpha \sin(\theta)}{\sqrt{2}}\;,
\end{align}
\end{subequations}
and requiring the $B_{\pi}$ component to be real. According to Eq.~(\ref{eq:rfBdef}), the $n^{\text{th}}$ RF field can then be written as
\begin{equation}
{\bf B}_{RF}(\omega_n t)=
{\bf B}_n
\cos(\omega_n t)\; ,
\end{equation}
where ${\bf B}_n$ is the RF field vector. It can be decomposed into its components along each Cartesian axis, ${\bf B}_n=\alpha_n \cos(\theta_n)\hat{{\bf e}}_{x}+\alpha_n \sin(\theta_n)\hat{{\bf e}}_{y}+B_{\pi_n}\hat{{\bf e}}_{z}$, and controlling the angle $\theta$ allows to chose orientation of the RF field. Considering $N$ such fields then leads to the Hamiltonian
\begin{equation}
\label{eq:FHamiltonian}
H({\bf r},t)=\mu_B g_F{\bf \hat{F}}\cdot \left({\bf B}_s({\bf r})+\sum_{n=1}^{N} {\bf B}_n \cos(\omega_n t)\right).
\end{equation}

We now move to a frame in which the direction of ${\bf B}_s({\bf r})$ always points along the $z$-axis by applying a rotation matrix $R$ to the Hamiltonian. While this can be done for any static field, here we assume that it is of Ioffe-Pitchard form (\ref{eq:IP}). In this case, the rotation matrix $R$ can be written as, 
\begin{equation}
  R=\begin{pmatrix}
      \frac{y}{r}&-\frac{x}{r}&0\\
       \frac{B_I x}{B_s r}&\frac{B_I y}{B_s r}&-\frac{G r}{B_s}\\
      \frac{G x}{B_s}&\frac{G y}{B_s}&\frac{B_I}{B_s}
 \end{pmatrix},
\end{equation}
and we define ${\bf B}'_n({\bf r})$ as the rotated RF field vector such that
\begin{equation}
  {\bf B}'_n({\bf r})=\begin{pmatrix}
  B'_{n_x}({\bf r})\\
  B'_{n_y}({\bf r})\\
  B'_{n_z}({\bf r})
  \end{pmatrix}
  = R\cdot{\bf B}_n \; .
\end{equation}
In the limit $|\mu_B g_F {\bf B}'_{n_z}({\bf r})| \ll \hbar \omega_{n}$ the component that oscillates parallel to the static
field, ${\bf B}'_{n_z}({\bf r})$, can be neglected and only the orthogonal parts, ${\bf B}'_{n_x}({\bf r})$ and ${\bf B}'_{n_y}({\bf r})$, contribute to the coupling between the atomic levels \cite{Hofferberth:07,Pegg:74}. The Hamiltonian (\ref{eq:FHamiltonian}) therefore becomes
\begin{align}
  H({\bf r},t)=\mu_B g_F \Bigg[& B_s({\bf r}) \hat{F}_z 
                +\left(\sum_{n=1}^N B'_{n_x}({\bf r})\cos(\omega_n t)\right) \hat{F}_x 
                \nonumber\\
  &+\left(\sum_{n=1}^N  B'_{n_y}({\bf r}) \cos(\omega_n t)\right) \hat{F}_y\Bigg] \; .
\label{eq:HamiltonianFinal}
\end{align}

\subsection{Single Frequency Floquet Matrix}
Let us briefly demonstrate the Floquet approach by first explicitly considering the situation where only a single frequency, $\omega_1$, is present. The Hamiltonian is then
periodic in time with a period of $\tau= \frac{2\pi}{\omega_1}$ and the
Floquet theorem states that the time-dependent Schr\"odinger equation
has a complete set of quasi-periodic solutions, which 
acquire a phase $\exp(-i \epsilon_n \tau)$ when $t\to t+\tau$. Here the $\epsilon_n$ are
quasi-eigenenergies and the phase factor defines $\epsilon_n$ mod
$\omega_1$.

The first step towards finding the quasi-eigenenergies is to remove
the time dependence of the Hamiltonian by replacing it with an
infinite matrix of the form
  \begin{equation}
  \begin{scriptsize}  
    H_F=\begin{pmatrix}
      .. &.. & .. & .. & ..& .. & ..\\
      .. &H_0+2\hbar\omega_1 & H_{-1} & H_{-2} & H_{-3} & H_{-4}&..\\
      .. &H_{1} & H_0+\hbar\omega_1 & H_{-1} & H_{-2} & H_{-3}&..\\
      .. &H_{2} & H_1 & H_{0} & H_{-1} & H_{-2}&..\\
      .. &H_{3} & H_2 & H_{1} & H_{0}-\hbar\omega_1 & H_{-1}&..\\
      .. &H_{4} & H_3 & H_{2} & H_{1}& H_{0}-2\hbar\omega_1 & ..\\
      .. &.. & .. & .. & ..& .. & ..\\
    \end{pmatrix},
  \end{scriptsize}  
  \end{equation}
where the {\it Floquet blocks} $H_n$ (with $n$ integer) are given by
\begin{equation}
  H_n({\bf r})=\frac{1}{\tau} \int_0^\tau \! H({\bf r},t)  e^{in\omega_1 t} \, dt\;.
\end{equation}

This procedure can be interpreted as an expansion of the original
Hamiltonian in terms of the Fourier components of $\omega_1$. For the Hamiltonian \eqref{eq:HamiltonianFinal} the Floquet blocks can be calculated by using the identity $\frac{1}{\tau} \int_0^\tau \! e^{i(m-n)\omega t} d t= \delta_{nm}$, where $\delta$ is the Kronecker delta \cite{Chu:04}. This leads to
\begin{widetext}
  \begin{equation}
    H_F= \begin{pmatrix}
      .. &.. & .. & ..&..& .. & ..\\
      .. &k B_s({\bf r})+2\hbar\omega_1 & 0 & 0 &\hbar\Omega^*_1({\bf r}) & 0 &..\\
      .. &0 & -k B_s({\bf r})+2\hbar\omega_1 & \hbar\Omega_1({\bf r}) &0 & 0& ..\\
      .. &0 & \hbar\Omega^*_1({\bf r}) & k B_s({\bf r})+\hbar\omega_1 &0& 0 & ..\\
      .. &\hbar\Omega_1({\bf r}) & 0 & 0 &-k B_s({\bf r})+\hbar\omega_1 & \hbar\Omega_1({\bf r}) & ..\\
      .. &0 & 0 & 0 &\hbar\Omega^*_1({\bf r}) & k B_s({\bf r}) & ..\\
      .. &0 & 0 & \hbar\Omega_1({\bf r}) &0& 0 & ..\\
      .. &.. & .. & ..&..& .. & ..\\
\label{eq:OurFloquetMatrixInf}
    \end{pmatrix}
  \end{equation}
where the Rabi frequency is given by $\Omega_1({\bf r})=\frac{\mu_B g_F}{4\hbar} \left(B'_{1_x}({\bf r})+i B'_{1_y}({\bf r})\right)$ and $k=\frac{\mu_B g_F}{2}$.

In order to obtain the eigenvalues and eigenvectors of $H_F$, it needs to be
truncated to a finite size, and a consistent way of doing this is to
fix the number of multiples of $\hbar\omega_1$ to include. This
corresponds to limiting the order of the photonic processes that can
occur, i.e.~if we limit the matrix to terms up to $ \pm 2\hbar
\omega_1$, maximally two photons can be absorbed and emitted. 
Including higher orders will give a more accurate description, but usually these terms are quickly decreasing in magnitude. To first order, this then leads to

  \begin{equation}
    H_F= \begin{pmatrix}
      k B_s({\bf r})+\hbar\omega_1 & 0 & 0 &\hbar\Omega^*_1({\bf r}) & 0 &0\\
      0 & -k  B_s({\bf r})+\hbar\omega_1 &\hbar \Omega_1({\bf r}) &0 & 0& 0 \\
      0 &\hbar \Omega^*_1 ({\bf r})& k B_s({\bf r}) &0& 0 &\hbar \Omega^*_1{\bf r} \\
      \hbar \Omega_1({\bf r}) & 0 & 0 &-k B_s({\bf r}) & \hbar \Omega_1{\bf r} & 0 \\
      0 & 0 & 0 &\hbar \Omega^*_1 ({\bf r})& k B_s({\bf r})-\hbar\omega_1 & 0  \\
      0 & 0 &\hbar \Omega_1({\bf r}) &0& 0 & -k B_s({\bf r})-\hbar\omega_1 \\
    \end{pmatrix},
  \end{equation}
  \end{widetext}
and diagonalising this matrix gives the quasi-energy spectrum,
which in turn allows the adiabatic RF potential to be calculated. This
process will be detailed in Sec.~\ref{sec:PotentialGen}.

\section{Multi Frequency Floquet Matrix}
In this section we will describe the treatment of an RF field containing several radio frequencies and use the model of Many Mode Floquet Theory (MMFT) \cite{Ho:83}. 
We will carry out the calculations explicitly for two different frequencies, $\omega_1$ and
$\omega_2$, with the extension to any number being straightforward. The Hamiltonian follows from
\eqref{eq:HamiltonianFinal} and is given by
\begin{small}
\begin{equation}
  H({\bf r},t)=\mu_B g_F{\bf \hat{F}}\cdot\left({\bf B}_s({\bf r})+{\bf B}_1({\bf r})\cos(\omega_1 t)+{\bf B}_2({\bf r})\cos(\omega_2 t)\right)\;,
\end{equation}
\end{small}
which leads to the basic form for the Floquet matrix
  \begin{scriptsize}  
  \begin{equation}
    H_F^{(2)}= \begin{pmatrix}
    .. &..& .. & .. &.. & .. &..\\
      .. &A+2\hbar\omega_r I& P & 0 &0 & 0 &..\\
      .. &P^T & A+\hbar\omega_r I& P &0 & 0& .. \\
      .. &0& P^T & A &P	& 0 & .. \\
      .. &0 & 0 & P^T & A-\omega_r I & P & ..\\
      .. &0& 0 & 0 &P^T & A-2\hbar\omega_r & ..  \\
      .. &..& .. & .. &.. & .. &..\\
    \end{pmatrix}.
  \end{equation}
\end{scriptsize} 
Here $I$ is the identity matrix and $\omega_r=\omega_1+\omega_2$ as we want to work in the dressed state picture. The elements of the $H_F^{(2)}$ matrix are matrices themselves and $H_F^{(2)}$ can therefore be thought of as a composite matrix of two single frequency Floquet matrices. The $\omega_1$ frequency terms and
the $\Omega_1({\bf r})$ Rabi frequency terms are contained in matrix $A$, which is
the same matrix as $H_F$ in the previous section.

The off diagonal elements $P$ contain the Rabi frequency of the second field, $\Omega_2({\bf r})=\frac{\mu_B g_F}{4\hbar} \left(B'_{2_x}({\bf r})+i B'_{2_y}({\bf r})\right)$, and are of the form
 \begin{equation}
P= \left(\begin{matrix}
..&.. & .. & .. & .. & .. & ..\\
..&0 & Y & 0 & 0 & 0 &..\\
..& 0 & 0 & Y & 0 & 0& ..\\
..&0 & 0 & 0 & Y & 0 & ..\\
..&0&0& 0 & 0 & Y & ..\\
..&0&0&0&0&0&..\\
..&..& ..&..&..&..&..\\
 \end{matrix}\right)\;,
\end{equation}
where
\begin{equation}
  Y= \begin{pmatrix}
    0 & \hbar\Omega_2^*({\bf r})\\
    \hbar\Omega_2({\bf r}) & 0\\
  \end{pmatrix}\;.
\end{equation}
\begin{widetext}
By following the method outlined above for the single frequency case, one can then evaluate each of these matrices. However, the explicit form of $H_F^{(2)}$ is too large to reproduce here and we only show a small excerpt of it below to highlight
the general structure. The key points of the structure are that the
bare trap eigenenergies are given along the diagonal and each Floquet
block differs by multiples of both $\omega_1$ and $\omega_r$. The
off-diagonal terms correspond to the coupling strengths and contain
terms for $\Omega_1({\bf r})$ and $\Omega_2({\bf r})$

  \begin{equation}
    H_F^{(2)}= \begin{pmatrix}
      k B_s({\bf r})+\hbar\omega_1+\hbar\omega_r & 0 & 0 &.. \\
      0 & -k B_s({\bf r})+\hbar\omega_1 +\hbar\omega_r& \hbar \Omega_1({\bf r})&..\\
      0 &\hbar \Omega_1^*({\bf r}) & k B_s({\bf r})+\hbar\omega_r& ..\\
      \hbar \Omega_1({\bf r}) & 0 & 0&.. \\
      0 & 0 & 0&..\\
      0 & 0 & \hbar \Omega_1({\bf r})&.. \\
      0 & 0 & 0&...\\
      0 & 0 & \hbar \Omega_2({\bf r}) &..\\
      0 & \hbar \Omega_2^*({\bf r}) & 0 &..\\
      \hbar \Omega_2({\bf r}) & 0 & 0&..\\
      0 & 0 & 0 &..\\
      0 & 0 & \hbar \Omega_2({\bf r}) &..\\
      .. & .. & ..& ..\\
    \end{pmatrix}.
\end{equation}
\end{widetext}
\section{Adiabatic Potential Generation}
\label{sec:PotentialGen}
In order to produce the adiabatic potential for the trapped atoms we need to extract the eigenvalues, $\epsilon_{z_n}^i$, and
eigenvectors, $\psi_{z_n}^i$, from the Floquet matrix for every
position on the spatial numerical grid, $z_n$. We do this numerically
as producing analytic expressions becomes quickly unfeasible.
\subsection{1D Potentials}
\begin{figure}[tb]
\centering
  \includegraphics[width=1\linewidth]{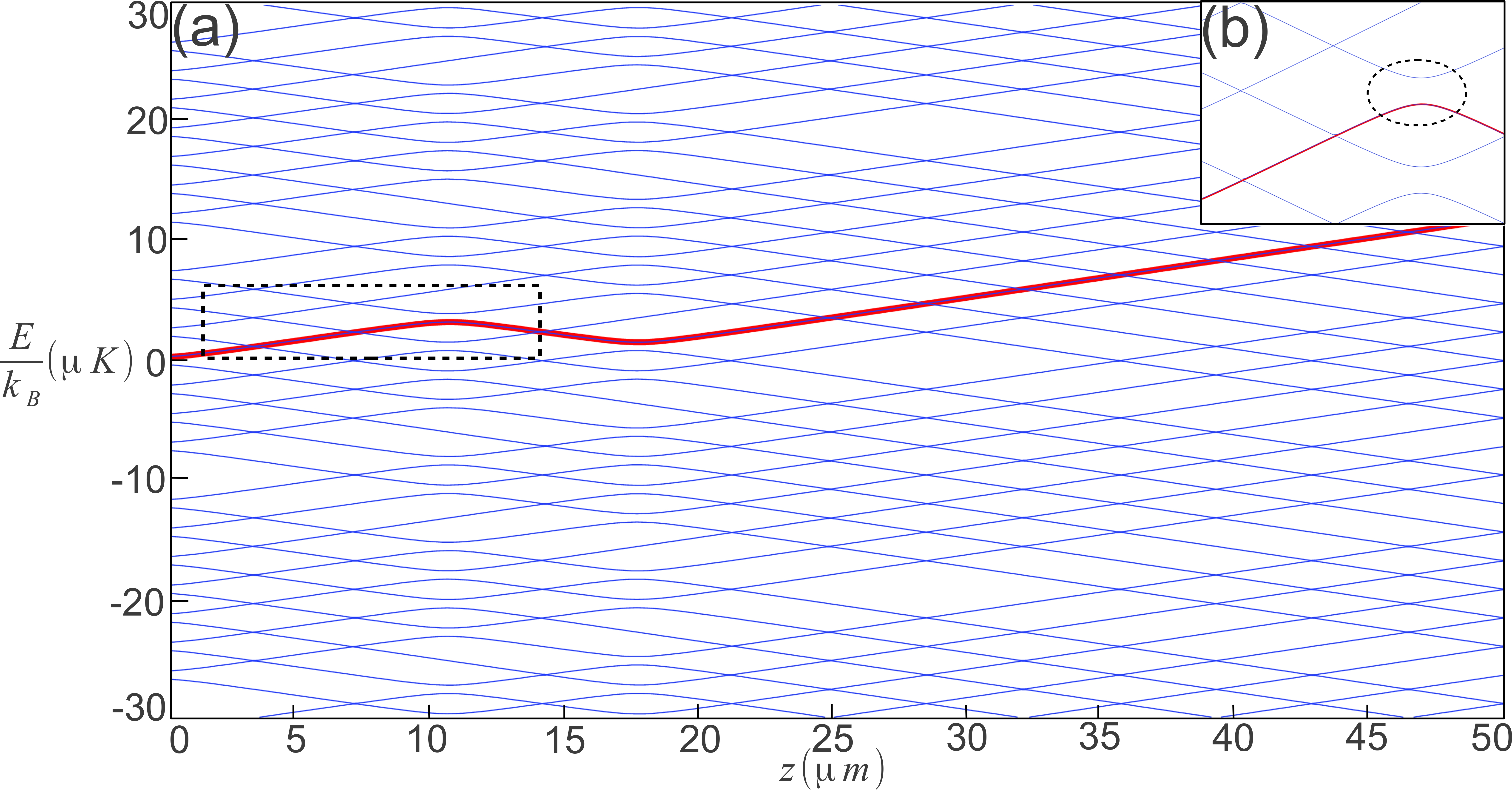}
  \caption{(Color online) (a) The spectrum of quasi eigenenergies for an atom 
    trapped in a IP trap irradiated by a two frequency, linearly
    polarised, RF field. The red (bold) line is the adiabatic RF
    potential produced from this spectrum. The inset (b) shows a zoom
    of the highlighted area in (a) where the avoided crossings can
    be seen (dotted circle). The RF fields have the frequencies $\omega_1= 2 \pi \times 1.5
    \times 10^{5}$Hz and $\omega_2= 2 \pi \times 4 \times 10^{5}$Hz
    with Rabi frequencies $\Omega_1=\Omega_2=9.274 \times 10^{4}$Hz. $m_F=\frac{1}{2}$ and $g_F=1$ in all cases.}
  \label{fig:Spectrum}
\end{figure}

The quasi-eigenenergies (the eigenvalues of $H_F$) generated when applying a two frequency, linearly polarised, RF field orthogonal to the movement of an atom in a one-dimensional IP trap are shown in Fig.~\ref{fig:Spectrum}. The
spectrum has a periodic structure, which stems from the fact that the
quasi eigenenenergies are defined mod $\omega_1$ and mod
$\omega_r$, and shows the expected avoided crossings (see
Fig.~\ref{fig:Spectrum}(b)). It should be noted that the Rabi frequencies $\Omega_1$ and $\Omega_2$ in this geometry are constant and can be changed independently from each other, which allows to control the size of the avoided crossings
associated with each frequency separately.

The potential seen by the atoms can now be found by identifying 
the energy eigenvalue in the centre of the trap which
is closest to the bare trap eigenenergy,  $\epsilon_{z_0}^A$ (normally the
smallest positive quasi-eigenvalue), and following its evolution in space. To find the
potential at the next numerical grid point $z_1$, we take the
eigenvector $\psi_{z_0}^A$ associated with the eigenvalue at
$\epsilon_{z_0}^A$ and calculate $\psi_{z_0}^A \cdot \psi_{z_1}^i$, the dot-product between it and all
eigenvectors corresponding to the eigenvalues $\epsilon_{z_1}^i$. The index of the eigenvector which
maximises this inner product gives the value of the potential at $z_1$. Iterating this process allows the construction of the potential along the whole length of the trap (see red line in Fig.~\ref{fig:Spectrum}).

\begin{figure}[tb]
\centering
  \includegraphics[width=1\linewidth]{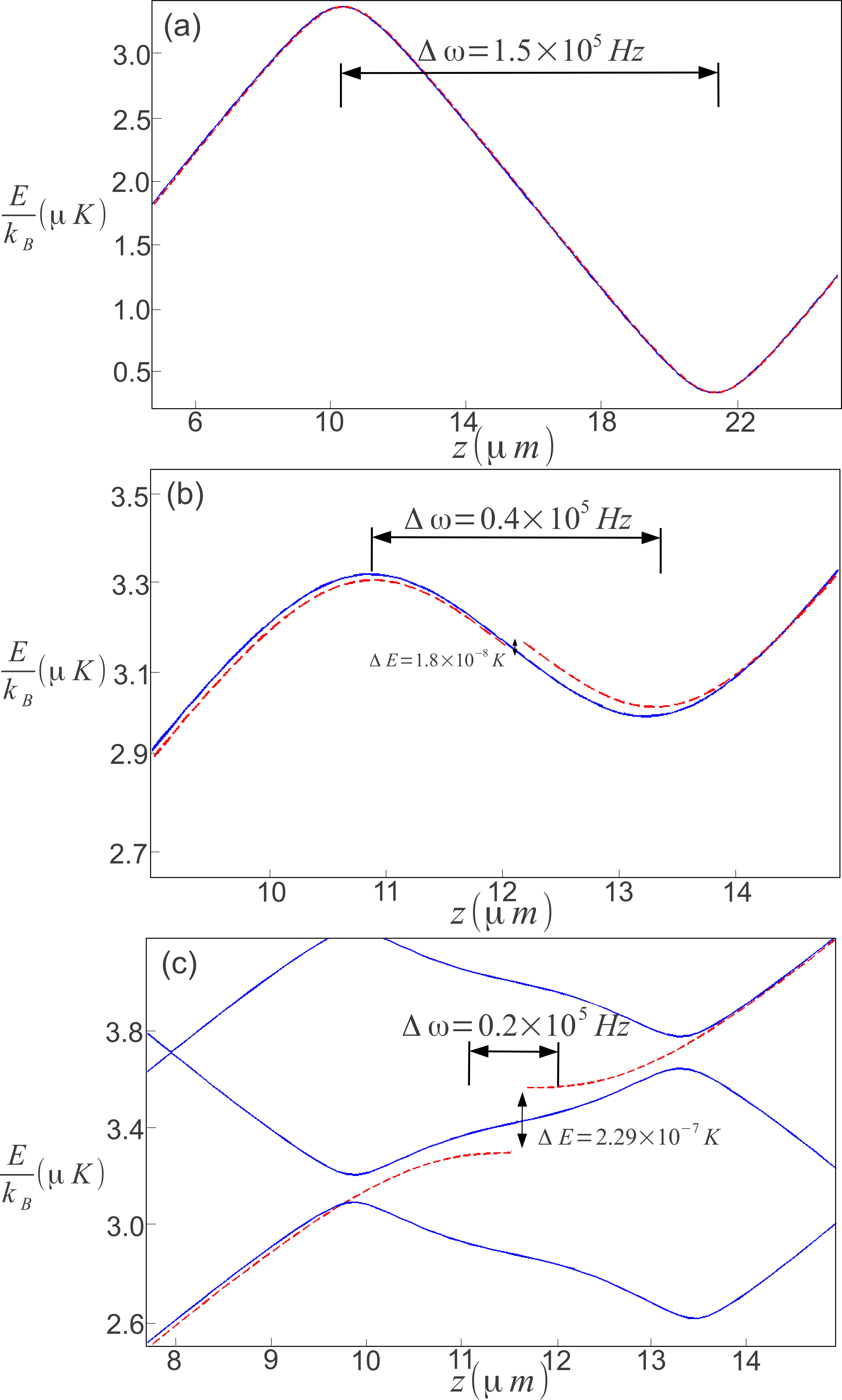}
  \caption{(Color online) Comparison between piecewise resonance model (Dashed Red Line) and the Floquet model (Solid Blue
    Line). All parameters are the same as for Fig~\ref{fig:PWPotential}. See text for discussion.}
  \label{fig:Comparison}
\end{figure}

Fig.~\ref{fig:Comparison} shows the adiabatic RF potentials generated using both the
piecewise resonance and Floquet methods for the same parameters as Fig.~\ref{fig:PWPotential}. 
One can see that the Floquet approach produces the same potential for (a) $\Delta \omega=1.5 \times 10^5 Hz$ and avoids the discontinuity for (b) $\Delta \omega=0.4 \times 10^5 Hz$, which is a notable improvement over the piecewise resonant model.

In the extreme case of $\Delta \omega=0.2 \times 10^5 Hz$ shown in Fig.~\ref{fig:Comparison} (c) the Floquet model produces another pair of avoided crossings on either side of the original avoided crossings seen at larger frequency spacing. These extra avoided crossings are the result of the frequencies no longer just coupling the local static field ${\bf B}_s({\bf r})$, where they are resonant. Instead, at this very close frequency separation, each frequency is applied to the effective field created by the static field and the other frequency, which leads to a shift in the resonance position and results in the additional avoided crossings. As one can also see from Fig.~\ref{fig:Comparison} (c), these avoided crossings are not present in the piecewise resonance model, since the rotating wave approximation eliminates the higher order coupling terms. 

From this example it is clear that the Floquet approach to modelling adiabatic RF potentials produces highly accurate results with significant corrections to the piecewise resonance approach. It is therefore well suited to accurately describe multi-frequency RF potentials created using closely spaced frequency. In the next section we will show how the Floquet approach can be extend to systems with multiple RF fields where the orientations of the fields are not necessarily perpendicular to the static field.

\subsection{2D Potentials}
\label{sec:2dPotential}
The well known flexibility of RF potentials stems from the large number of parameters (number of frequencies, frequency spacing, field strength and orientation) that can be controlled to modify the geometry of the potentials. These parameters can also be time-dependent, which allows to use the resulting potentials to instigate adiabatic or non-adiabatic dynamics. In the following we consider a multi-dimensional setup and use the Floquet approach to describe some of the features which appear when taking these additional parameters into account.

The main change necessary to extend the model to higher dimensions is to use a multi-dimensional static trapping field $B_s({\bf r})$ which then leads to coupling terms $\Omega({\bf r})$ of the same, higher dimension. This means that the Floquet matrix contains elements that are multidimensional, but its basic structure is unchanged from the one-dimensional case. The added dimension however means that the numerical processing power needed to diagonalise the Floquet matrix is significantly higher.

In the following we will show an example of an interesting potential that can be created using just two frequencies. The numerical approach is similar to the one detailed above, in that we construct the 2D potentials by splitting them into 1D slices and combine them to produce the full 2D potential. To ensure that our minimum rotation method produces 1D slices which are consistent with each other, we use the previously generated 1D slice for each consecutive 1D slice to define an initial quasi-eigenvector and eigenvalue pair.

\subsubsection{Two Frequency Potential}
\begin{figure}[tb]
\centering
  \includegraphics[width=1\linewidth]{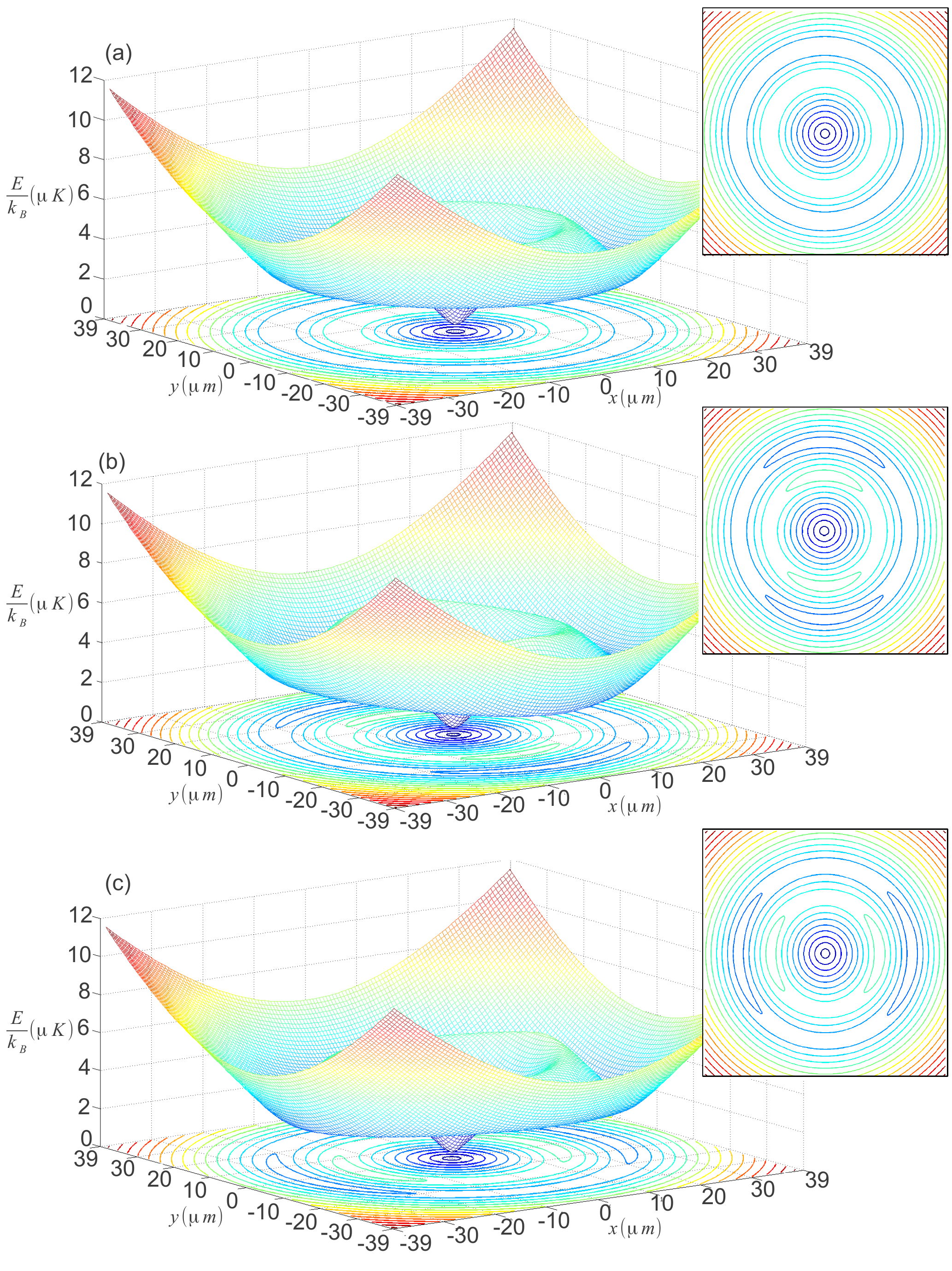}
  \caption{(Color online) The adiabatic potential of an atom trapped in a IP trap irradiated by 
  two RF fields with frequencies of $\omega_1=3 \times 10^{5}$Hz and $\omega_2=4\times10^{5}$Hz for (a) $\alpha=0$, $B_{\pi}=8 \mu \text{T}$ (b) $\alpha=10\mu \text{T}$, $\theta=0$, $B_{\pi}=2 \mu \text{T}$ and (c) $\alpha=10 \mu \text{T}$, $\theta=\pi/2$, $B_{\pi}=2 \mu \text{T}$. In all cases $m_F=\frac{1}{2}$ and $g_F=1$ .}
  \label{fig:2DPotential}
\end{figure}

Two-dimensional RF potentials generated using one frequency have already been demonstrated to study rotating BECs in annular geometries \cite{Morizot:06}. One of the simplest ways to create a more involved structure using two frequencies is shown in Fig.~\ref{fig:2DPotential} (a) where we consider a 2D IP trap and a constant RF field vector which contains a $B_{\pi}$ component only ($\alpha=0$). One can see that the geometry one can create consists of an outer ring and an inner potential, which is actually simply the result from the one-dimensional model of Sec.~\ref{sec:PotentialGen} rotated around the axis along z=0.

However, by controlling $\omega_1$ and $\omega_2$ one can easily change the depth and radius of the inner harmonic well and the outer ring, as well as the size and height of the barrier between them. Performing such a change in a time-dependent way one can then engineer a situation where the tunnel coupling between the ring and inner trap can be increased or decreased with high accuracy. 

Furthermore, having control over the orientation of the RF fields allows one to break the rotational symmetry and create potentials that are no longer isotropic. In Fig.~\ref{fig:2DPotential} (b) and (c) we show two examples of a 2D potential generated by using constant RF field vectors for (b) $\alpha=10\mu \text{T}$, $\theta=0$, $B_{\pi}=2 \mu \text{T}$ and (c) $\alpha=10 \mu \text{T}$, $\theta=\pi/2$, $B_{\pi}=2 \mu \text{T}$. The resulting potentials display two minima and two maxima on the rings of resonance, which can be understood by considering that the orientation of the $B_{n_x}\hat{{\bf e}}_x$ and $B_{n_y}\hat{{\bf e}}_y$ components of the RF field with respect to the static field is not constant in space. This leads to the variable coupling. Furthermore, controlling the angle $\theta$ allows to create RF fields which point in any direction, which in turn allow the creation of minima and maxima at any position and even in a time dependent fashion.

\subsubsection{Ring Vortex Solitons}
\label{sec:RVS}
As an example of how useful the time-dependent control of the tunnelling interaction between the outer ring and the inner harmonic trap can be, we demonstrate in this section how the potential discussed above can be used to create so-called ring vortex solitions (RVSs) in gaseous Bose-Einstein condensates. These states consist of multiple concentric density-wave rings with a non-zero winding number and they were recently extensively studied by Li {\it et al.} \cite{Li:12}. To create them in a laboratory, one can consider starting out with a single-frequency RF field in a two-dimensional setting, and creating a condensate carrying a persistent current in the resulting annular potential. Adding the second frequency would allow one to create the central  potential and by adjusting all frequencies and intensities appropriately, the current would start to tunnel into the inner potential. 

To simulate the creation of RVSs in such a way, the potential must allow large tunnelling strength and therefore the model must be able to deal with situations where two frequencies come close. This makes the Floquet approach the most suitable option.

\begin{figure}[tb]
\centering
  \includegraphics[width=1\linewidth]{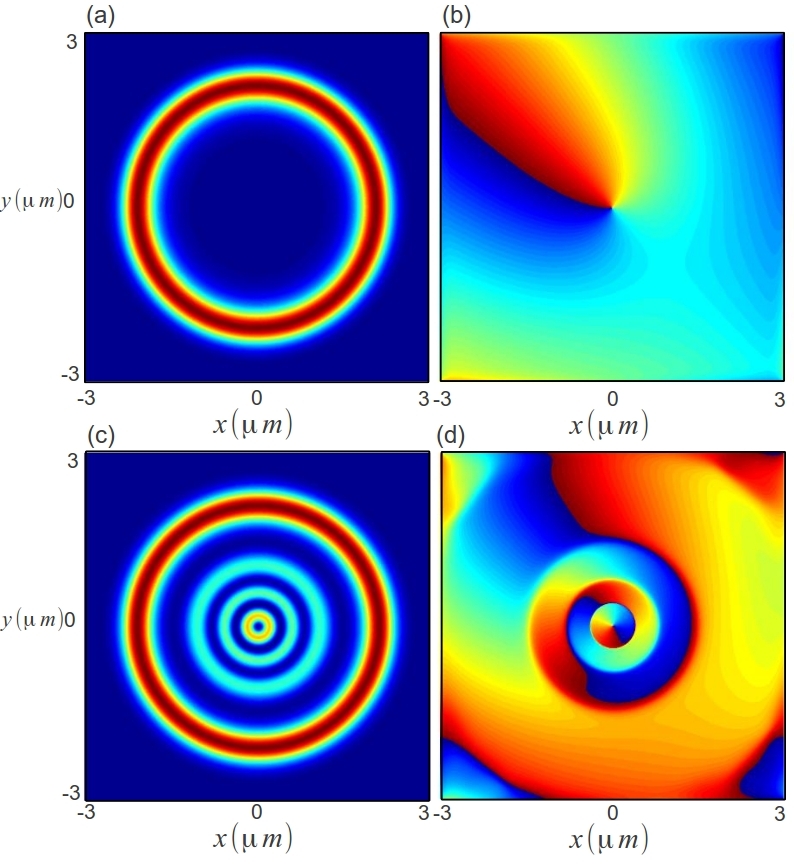}
  \caption{(Color online) The creation of a RVS in our adiabatic RF potential. The adiabatic RF potential is created by irradiating an IP trap with two RF fields of frequency $\omega_1=27 \times 10^{3}$Hz and $\omega_2=33\times10^{3}$Hz with $ B_{\pi}=0.4219 \mu \text{T}$ for both fields. (a) Shows the density of the BEC of winding number n=1 initially confined to the outer ring and (b) is its corresponding phase plot. (c) Shows the density of the BEC at $t=800 \mu s$ and (d) is corresponding phase plot.}
  \label{fig:RVS} 
\end{figure}

In Fig.~\ref{fig:RVS} we show a simulations of this tunnelling process using a Gross-Pitaevskii model for the condensate and a potential calculated using the Floquet approach outlined above. The density and phase distribution for the BEC initially confined to the outer ring with $n=1$ can be seen in Figs.~\ref{fig:RVS} (a) and (b), and the density and phase after transfer into the harmonic trap is shown in Figs.~\ref{fig:RVS} (c) and (d). Clearly the resulting state can be identified as a ring vortex soliton.

\subsubsection{Concentric Ring Potentials}
While we have explicitly shown a 2D adiabatic RF potential generated using two frequencies, this can easily be extended to include more frequencies to create potentials that consist of multiple concentric rings. As in the previous example of just two frequencies, control over the frequency spacing would allow one to adjust the tunnel coupling between the individual rings and can be accurately determined by using a Floquet approach. Such systems are highly desirable as they offer experimentally realistic systems in which one could observe rotational fluxons (Josephson vortices) \cite{Brand:09} or study the Kibble-Zurek mechanism in BECs \cite{Brand:13}.
\section{Conclusion}
\label{sec:Conclusion}
We have shown that using a non-perturbative Floquet approach to calculate radio-frequency potentials for atoms on atom chips leads to good results for multi-frequency situations in one and two dimensions. This approach offers significant corrections to a piecewise-resonant model and can be used to accurately describe situations where two frequencies come close (for example when trying to engineer large tunnel couplings). We have also shown that a Floquet approach can deal with higher dimensions and RF fields of arbitrary field orientation. We have demonstrated the flexibility of the Floquet approach by considering the creation of ring vortex solitons in a 2D adiabatic RF potential which requires that the tunnelling dynamics be determined accurately. Such potentials can be used in a variety of ways, and for example, allow the creation of rotational fluxons and non standard trapping potentials that can connect different topological geometries. This will be the focus of future work. 

\section{Acknowledgements} 
The authors would like to thank Igor Lesanovsky for valuable discussions. This work was supported by the European Science Foundation (ESF) 
within the framework of the common perspectives for cold atoms, semiconductor polaritons and nanoscience activity and Science Foundation Ireland under project numbers 05/IN/I852 and 10/IN.1/I2979.
  

\begin{thebibliography}
{}

\bibitem{Folman:02} R.~Folman, P.~Kr\"uger, J.~Schmiedmayer, J.~Denschlag and C.~Henkel, Adv.~At.~Mol.~Opt.~Phys. {\bf 48}, 263 (2002)

\bibitem{Fortagh:07} J.~Fort\'{a}gh and C.~Zimmermann Rev.~ Mod.~ Phys.~{\bf 79},
  235 (2007)

\bibitem{Schumm:05} T.~Schumm, S.~Hofferberth, L.M.~Andersson,
  S.~Wildermuth, S.~Groth, I.~Bar-Joseph, J.~Schmiedmayer and
  P.~Kr\"uger, Nature Physics {\bf 1}, 57 (2005)

\bibitem{Lesanovsky:06} I.~Lesanovsky, T.~Schumm, S.~Hofferberth,
  L.~M.~Andersson, P.~Kr\"uger, and J.~Schmiedmayer.
  Phys.~Rev.~A. {\bf 73}, 033619 (2006)
  
\bibitem{Fernholz:07} T.~Fernholz, R.~Gerritsma, P.~Kr\"uger,
  R.J.C.~Spreeuw, Phys.~Rev.~A {\bf 75}, 063406 (2007)

  \bibitem{Morizot:06} O.~Morizot, Y.~Colombe, V.~Lorent, H.~Perrin, 
  and B.M.~Garraway, Phys.~Rev.~A {\bf 74}, 023617 (2006)
  
\bibitem{Hofferberth:07} S.~Hofferberth, B.~Fischer, T.~Schumm,
  J.~Schmiedmayer, and I.~Lesanovsky, Phys.~Rev.~A {\bf 76}, 013401
  (2007)

\bibitem{Sinuco:12} G.~Sinuco-Le\'on and B.~M.~Garraway, New J.~Phys. {\bf 14} 123008 (2012)
  
\bibitem{Courteille:06} Ph.W.~Courteille, B.~Deh, J.~Fort\'agh,
  A.~G\"unther, S.~Kraft, C.~Marzok, S.~Slama, C.~Zimmermann,
  J.~Phys.~B: At.~Mol.~Opt.~Phys.~{\bf 39}, 1055 (2006)
  
\bibitem{Morgan:11} T~.Morgan, B.~O’Sullivan, and Th.~Busch, Phys.~Rev.~A {\bf 83}, 053620 (2011)      

\bibitem{Shirley:63} J.~H.~Shirley, Phys.~Rev.~ {\bf 138}, B979 (1965)

\bibitem{Ho:83} T.~S.~Ho, S.~I.~Chu, and J.~V.~Tietz,  Chem.~Phys.~Lett.~ {\bf 96}, 464–471 (1983) 

\bibitem{Ho:84} T.~S.~Ho and S.~I.~Chu, J.~Phys.~B: At.~Mol.~Phys. {\bf 17}, 2101–2128 (1984) 

\bibitem{Drese:99} K.~Drese and M.~Holthaus Eur.~Phys.~J.~D {\bf 5}, 119–134 (1999)

\bibitem{Son:08} S.~K.~Son and Shih-I.~Chu Phys. Rev. A {\bf 77}, 063406 (2008) 

\bibitem{Chu:04} S.~I.~Chu, D.~A.~Telnovc, Physics Reports, {\bf 390}, 1-131 (2004)

\bibitem{Tannoudji:92} C.~Cohen-Tannoudji, J.~Dupont-Roc, and G.~Grynberg, {\it Atom-Photon Interactions} (Wiley, New York, 1992)

\bibitem{Bloch:40} F.~Bloch and A.~J.~F. Siegert, Phys.~Rev. {\bf 57}, 522 (1940) 

\bibitem{Pegg:74} D.~T.~Pegg, J.~Phys.~B {\bf 6} 241 (1974)
\bibitem{Li:12} J.~Li, D.S.~Wang, Z.Y.~Wu, Y.M.~Yu and W.M.~Liu, Phys.~Rev.~A {\bf 86} 023628 (2012)
\bibitem{Brand:09} J.~Brand, T.~J.~Haigh and U.~Z\"ulicke, Phys.~Rev.~A {\bf 80} 011602 (2009)
\bibitem{Brand:13} Shih-Wei Su, Shih-Chuan Gou, A.~Bradley, O.~Fialko and J.~Brand, Phys.~Rev.~Lett.~{\bf 110} 215302 (2013)
\end{thebibliography}
\end{document}